\documentclass[hidelinks]{elsarticle}

\makeatletter
\def\ps@pprintTitle{%
	\let\@oddhead\@empty
	\let\@evenhead\@empty
	\def\@oddfoot{\centerline{\thepage}}%
	\let\@evenfoot\@oddfoot}
\makeatother

\usepackage{graphicx}
\usepackage{sidecap}

\usepackage[margin=1.2in]{geometry}
\usepackage{hyperref}
\usepackage[labelfont=bf]{caption}

\begin{document}
	\title{Measuring and modeling interventions in aging}
	\author[1,2]{Nicholas Stroustrup}
	\ead{nstroustrup@post.harvard.edu}
	\address[1]{Centre for Genomic Regulation (CRG)\\The Barcelona Institute of Science and Technology\\Dr. Aiguader 88, Barcelona 08003, Spain}
	\address[2]{Universitat Pompeu Fabra (UPF), Barcelona, Spain }
	
	\begin{abstract}
At the physiological level, aging is neither rigid nor unchangeable. Instead, the molecular and mechanisms driving aging are sufficiently plastic that a variety of diverse interventions--dietary, pharmaceutical, and genetic--have been developed to radically manipulate aging. These interventions, shown to increase the health and lifespan of laboratory animals, are now being explored for therapeutic applications in humans.

This clinical potential makes it especially important to understand how, quantitatively, aging is altered by lifespan-extending interventions.  Do interventions delay the onset of aging? Slow it down? Ameliorate only its symptoms?  Perhaps some interventions will alter only a subset of aging mechanisms, leading to complex and unintuitive systemic outcomes.  Statistical and analytic models provide a crucial framework in which to interpret the physiological responses to interventions in aging.  

This review covers a range of quantitative models of lifespan data and places them in the context of recent experimental work.  The careful application of these models can help experimentalists move beyond merely identifying statistically significant differences in lifespan, to instead use lifespan data as a versatile means for probing the underlying physiological dynamics of aging.
\end{abstract}

\maketitle
\begin{itemize}
\item Death involves the final collapse of vital physiological networks, and the timing of this collapse provides a systems-level measure of aging.   
\item Many of the best models for lifespan data common in the clinical literature are rarely applied in basic research studies of aging.
\item Multivariate regression models allow differences between experimental replicates to be explicitly measured and accounted for when estimating the effect of interventions.
\item Semi-parametric models allow interventions to be studied with fewer implicit assumptions regarding the empiric data.
\item Frailty models extend parametric and semi-parametric approaches to more accurately account for unmeasured heterogeneity.
\item Competing risk models and mixture models provide a formal framework for studying multi-causal aging processes.
\item Common models, like the Gompertz parametric model, are accurate only when experimental data exhibit particular geometric patterns.  These patterns are often not present.
\end{itemize}

\begin{figure}[h!]
\centering
\includegraphics[width=.85\textwidth]{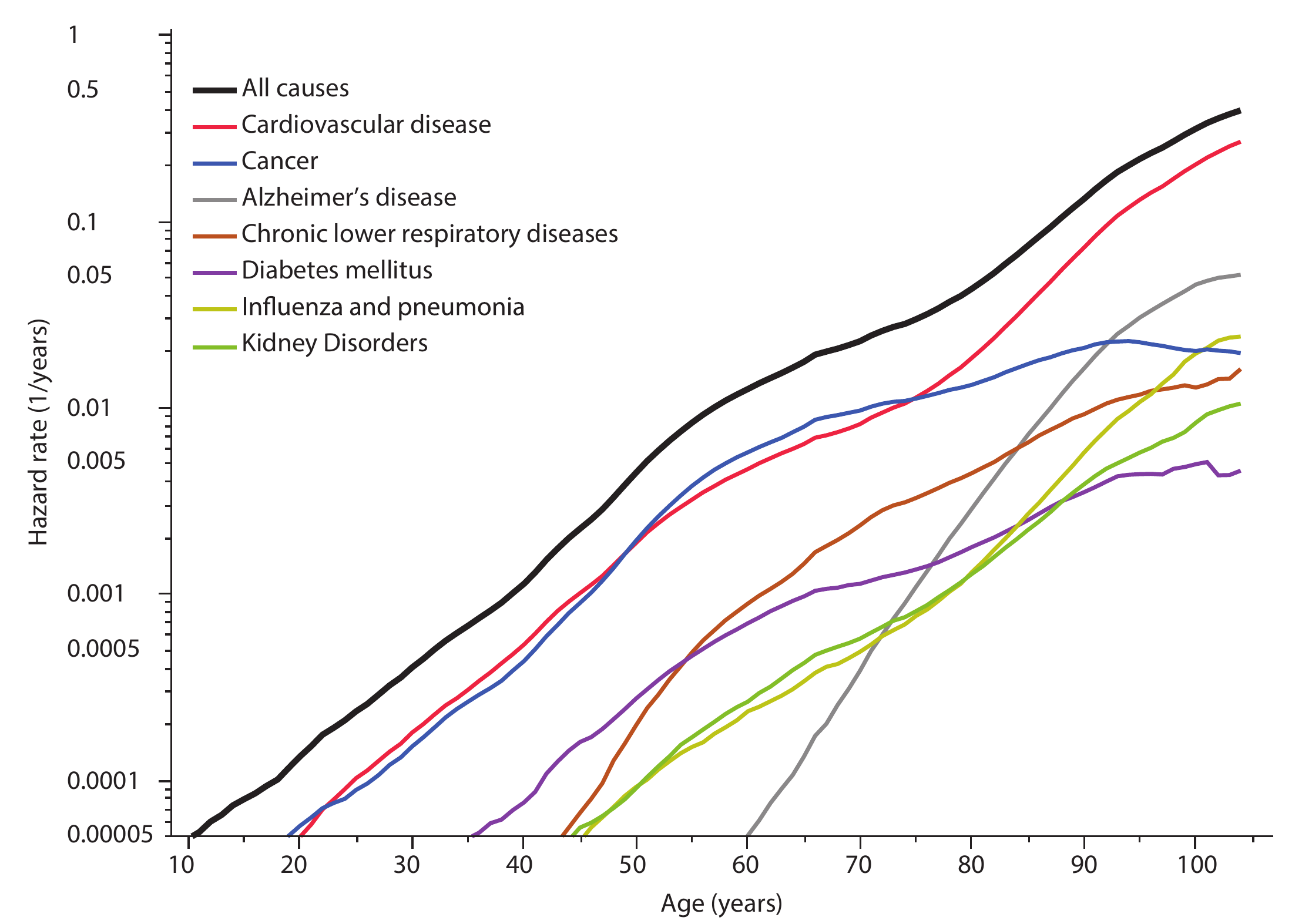}
\caption{\textbf{Systems-level measurement of complex physiological processes.}  The risk of death from the seven most frequent causes of non-accidental death is shown, corresponding to 70\% of all deaths reported in the USA in 2015 \cite{murphy2015death_data}. Cause-specific risks \emph{(colored lines)} sum up to produce the all-causes hazard function \emph{(black)}.  Each cause exhibits distinctive age-dependent effects, though substantial correlations exist among the causes.   }
\label{fig_1} 
\end{figure}

\newpage
Douglas Adams once said that ``There is an art to flying, or rather a knack. The knack lies in learning how to throw yourself at the ground and miss."  Immortality requires learning a similar knack: you must first be born and then subsequently avoid dying forever.  This is challenging in part because there are many different causes of death to avoid--accidents, infections, cancer, heart disease, neurodegenerative disorders--and because our bodies slowly change in ways that make most of these causes of death increasingly probable (Fig. \ref{fig_1}).  As the risks of occurrence for different diseases increase, their contributions add up to produce a doubling in all-cause mortality risk approximately every eight years \cite{murphy2015death_data}.  The specific physiological changes driving these increases in disease-specific mortality risk remain uncertain, motivating intense research into the molecular, cellular, and systems biology of aging.  

The mathematics governing increases in the risk of death are non-trivial and can appear obscure or arcane to experimental biologists who may lack formal training in probability theory and statistics.  Yet a general understanding of the models used to analyze lifespan data are nevertheless important for molecular, cell, and systems biologists.  In a clinical context, the mathematics and statistical models of lifespan highlight the many challenges ahead for proponents of radical lifespan extension--delaying or eliminating a small number of causes of death can yield only an incremental effect on lifespan. For example, it is estimated that curing all cancers would produce less than four years of extended lifespan\cite{mackenbach1999gains}.  The reason for this is intuitive--elderly patients, cured of a particular disease, remain highly likely to succumb to any of multiple other diseases, and so live (on average) only a little longer--but explanations of the statistics governing this are often impenetrable for non-specialists. A basic understanding of quantitative models are also important in the basic research, as much of the work in molecular and cell biology today remains focused on mechanisms whose involvement in aging was at least initially justified by those mechanisms' effect on lifespan. Experimental researchers need to be critical judges of this evidence.  

\section{Describing aging using hazard and survival functions}
Most analyses of lifespan data are grounded in study of two related mathematical functions--the survival curve and the hazard function.  The hazard function provides an intuitive measure of the risk of death, defined as the probability that a typical individual who is currently alive will soon die.  This probability is much higher in older individuals compared to younger ones, usually interpreted as evidence of some physiological weakness or susceptibility to death shared among old individuals not present in the young.  Over the last forty years, it has become very clear that this increase in risk does not emerge from some universal natural law. The shape of hazard functions is a product of evolutionary forces and varies enormously between species \cite{jones2014diversity}.  In fact, a mammal, the naked mole rat, has recently been shown to exhibit a nearly constant hazard function \cite{ruby2018naked}.  Formally, the hazard function is defined as a conditional probability, $h(t) = \lim P(T \le t+\Delta t| T > t)$ as  $\Delta t \rightarrow 0 $, and is usually estimated and plotted as the rate $h(t) = P(t< T \le t+\Delta t) / P(T > t)$ .

The survival curve is a different function that describes the fraction of a population remaining alive over time.   At the start of an observational period, this fraction is one and then drops each time an individual dies.  The survival function is formally defined as the cumulative probability of remaining alive, $S(t) = P(T > t)$.  The cumulative probability of remaining alive obviously is related to the risk of death, governed by the relation $S(t) = \exp \left( - \int_0 ^{\tau} h\left( \tau \right) d \tau \right)$ . Though the hazard function often provides a clearer visualization of patterns in mortality, any model of lifespan data can be equivalently stated in terms of the survival function.

\section{Identifying changes in lifespan with non-parametric methods}
The analysis of lifespan data usually involves the application of a non-parametric test used to identify statistically significant changes in lifespan.  Common methods include the log-rank, Wilcox, and the modified Kolmogorov-Smirnov (KS) tests, all of which ask whether two populations’ lifespan correspond to the same underlying survival and hazard function.  These tests make relatively few assumptions about the statistical properties of the underlying lifespan data, and so have remained in continuous use for decades without substantial modification.  However, this lack of assumptions leads to a lack of interpretability--most non-parametric tests can show that lifespan has been altered but not how it has been altered.  Recently, non-parametric approaches have been developed to distinguish changes in mean lifespan from changes in the variation in lifespan, as part of a pace-shape framework \cite{john2018diet}.

\section{Modeling the hazard function with parametric models}
\begin{figure}
\centering
\includegraphics[width=0.85\textwidth]{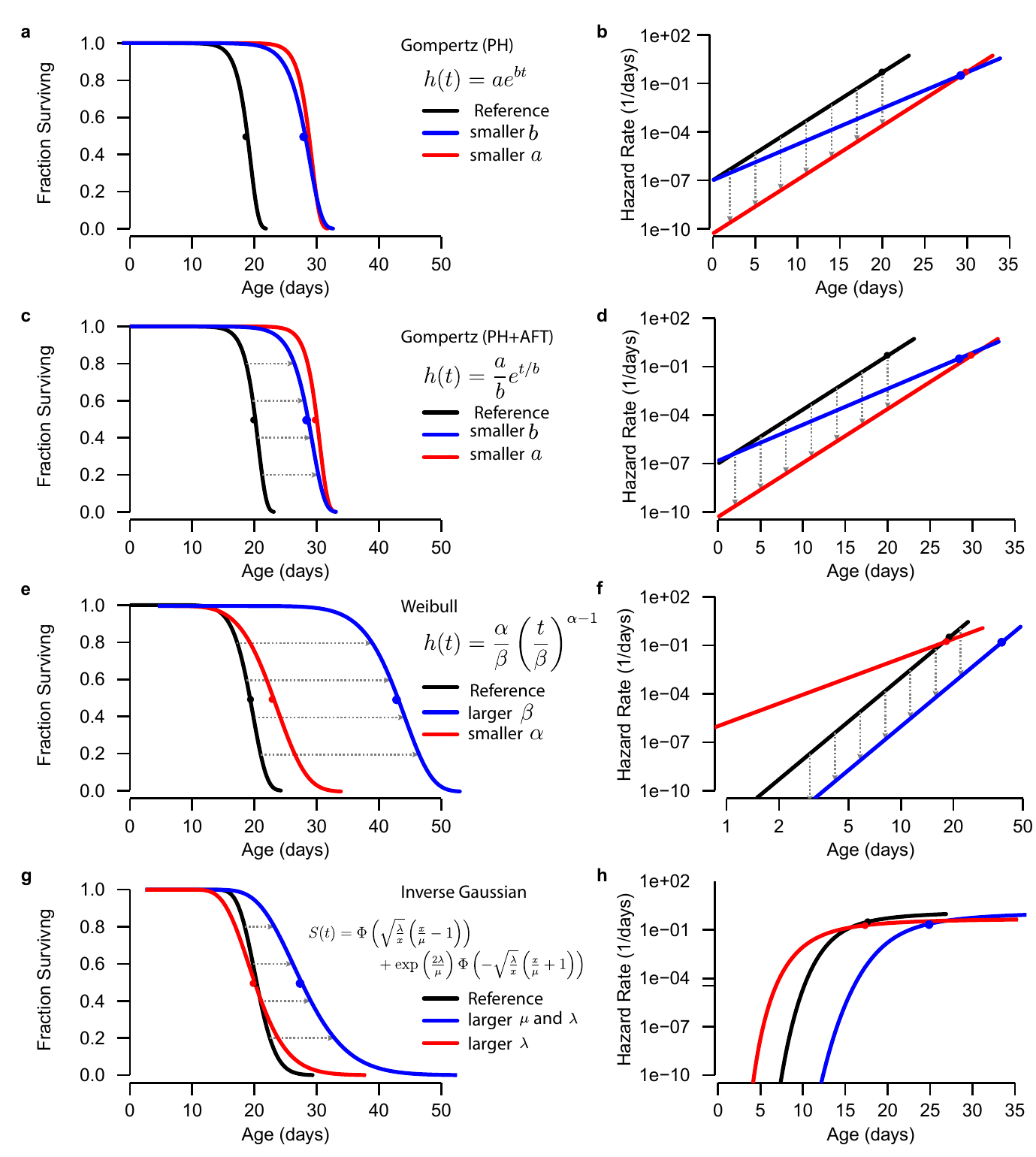}
\caption{\textbf{Parametric models of lifespan data.}  Various simple functions have been proposed to approximate empiric lifespan data, shown as models of survival functions \emph{(left)} and \emph{(right)} hazard functions.  Median lifespan is marked as a single point on each corresponding curve.  Geometric regularities in the influence of various parameters are marked with gray arrows.  \textbf{a-b. } The Gompertz distribution is commonly employed with the parameter $a$ that determines the risk of death in animals of zero age and $b$ that determines the rate of increase in risk over time.  \textbf{c-d. } The alternate parameterization of the Gompertz model removes the implicit and often missed time-scale dependence of the $a$ parameter, allowing changes in initial mortality and changes in doubling time to be isolated, and allowing the Gompertz function to model both changes in proportional hazards and changes in timescale. \textbf{e-f.} The Weibull distribution models hazard functions that increase as a polynomial of time (in contrast to the exponential increase assumed in by the Gompertz model).  Weibull hazard functions therefore form straight lines when plotted on log-log axes, rather than log-linear axes. \textbf{g-h.} Inverse Gaussian distributions exhibit inherently decelerating hazard functions, and provide a link between lifespan data to the theory of Weiner Processes and Brownian motion. }
\label{fig_2} 
\end{figure}

Parametric models go beyond non-parametric models by positing that lifespan data can be accurately described by simple mathematic functions with a small number of free parameters.  In data where this assumption holds,  parametric functions allow researchers to describe complex data sets using a small number of parameters, providing straightforward geometric interpretations of how interventions effect lifespan.  For example, the commonly used Gompertz model assumes that populations exhibit hazard functions that increase exponentially over time (Fig. \ref{fig_2} a-d), following a straight line on a log-linear plot.  This model has been shown to provide an adequate approximation for some human populations \cite{pakin1984critical,vaupel1998biodemographic} as well as some invertebrate populations \cite{ tatar1993long, mair:2003}.  Yet, the shape of hazard functions vary enormously between species \cite{jones2014diversity} and despite its popularity, the Gompertz model is frequently out-performed by several other two-parameter distributions.  These include the Weibull model\cite{rinne2008weibull}, that assumes a polynomial increase in the risk of death over time (Fig. \ref{fig_2} e-f), as well as the inverse-Gaussian model (Fig. \ref{fig_2} g-h) which notably has a compelling theoretic grounding in the statistical physics of Brownian motion \cite{weitz:2001vl,aalen:2001}.  Other parametric alternatives include Gompertz-Makeham, log-logistic, and log-normal models.  Parameter estimates can be obtained by a variety of methods\cite{lenart2012gompertz,de2018reassessment}, with maximum-likelihood estimation approaches\cite{flexsurv2016} almost always producing the most accurate results.

Different explanations have been proposed to explain why physiological aging processes might yield lifespan distributions conforming to simple parametric forms.  The more quantitative of these theories draw on a mix of reliability theory \cite{gavrilov:2001}, complex networks theory \cite{vural:2014}, and statistical physics \cite{strehler:1960}, and this area remains a topic of active research.  One interpretation shared among most parameterizations involves a single “timescale” or  “rate” parameter that uniquely governs the time-dependent increase of the hazard functions including the Gompertz $b$ parameter, the Weibull $\beta$ parameter, and the Inverse Gaussian $\lambda$ parameter.  These timescale parameters are used to quantify a population's ``rate of aging”, a measure of some average speed at which physiologically young individuals change into old individuals.  This rate of aging is then distinguished from other model parameters that describe distinct aspects of aging distinct from essential rate, allowing interventions to be qualitatively categorized according to their distinct effects on different parameters \cite{wu2009multiple, giannakou2007dynamics, mair:2003, ruvkun:2007}, \cite{hughes2016different}.  Most of the common parametric models differ primarily in the interpretation of their non-timescale parameter and secondarily in the accuracy of their approximation of empiric data.  
The major limitation to parametric methods is that for most data sets, there does not exist a single unambiguously best parameterization for the underlying empiric data.  In cases where different parameterizations can equally-well approximate empiric data, the different parametrizations provide multiple, discordant interpretations of the same underlying data. The most common reasons for this are described later in the “frailty” section. 

\section{Modeling changes in lifespan with semi-parametric methods}
Semi-parametric models improve on parametric approaches by eliminating the necessity of picking one simple mathematic function to represent survival and hazard functions.  Semi-parametric models parameterize only the action of the intervention itself--hence the name ''semi-parametric".  They provide a more flexible means  for making relative statements comparing two populations.  Theory would predict that semi-parametric methods require larger population sizes compared to parametric methods to achieve the same statistical power and significance.  However these costs are often minor \cite{bokov2017biologically}, especially compared to the complications that arise when poor-fitting fully-parametric forms are used.

\begin{SCfigure} 
\centering
\includegraphics[width=0.6\textwidth]{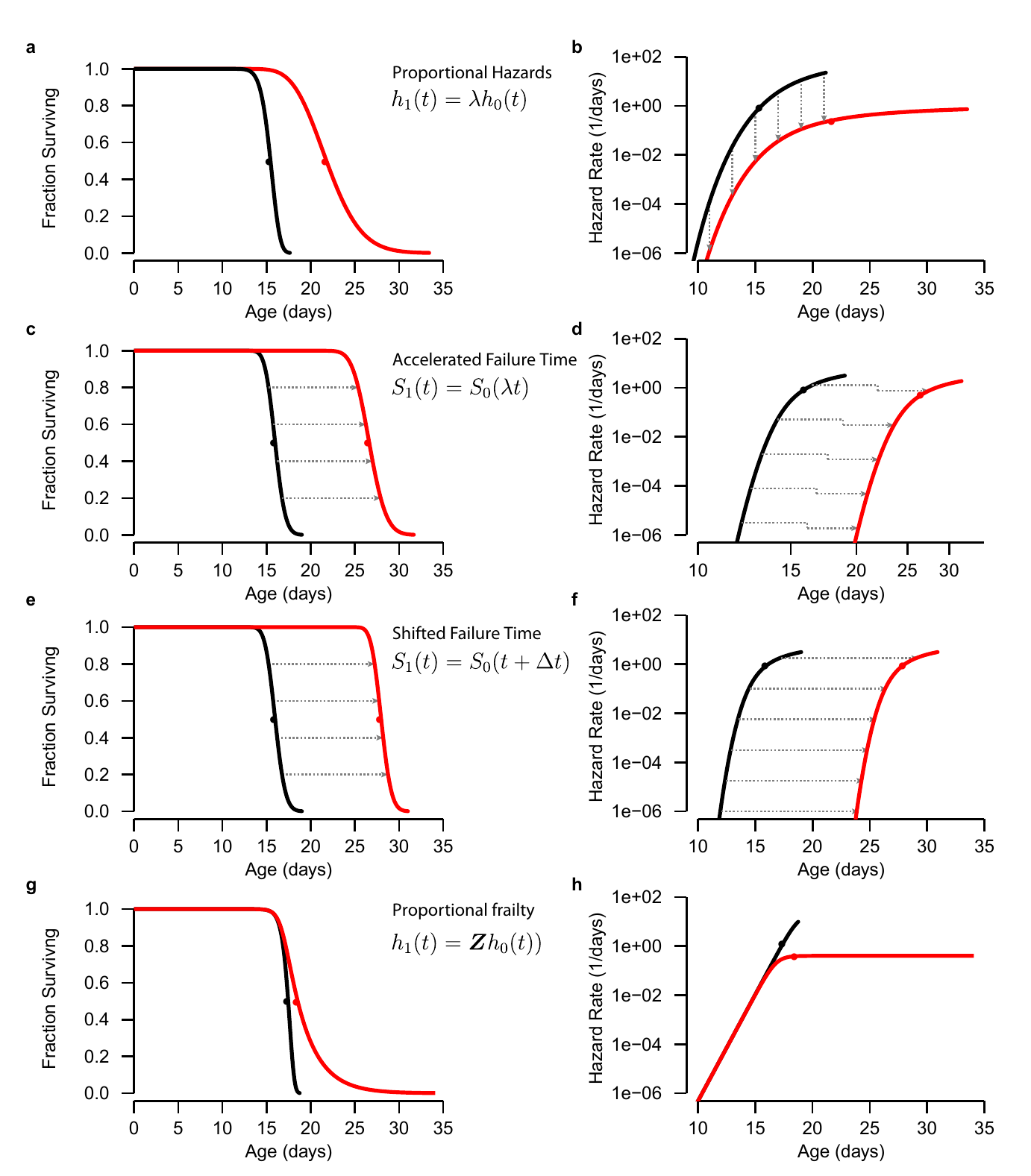}
\caption{\textbf{Semi-parametric models of lifespan data. }  Semi-parametric models describe differences between populations in a way that does not depend on any particular parametric form of the survival curve of hazard function. \textbf{a-b.} Proportional hazards functions assume that two populations' hazard functions are offset by a constant ratio. \textbf{c-d.} Accelerated Failure time models assume that two populations' survival curves are related by a temporal scaling, corresponding to simultaneous shift of the hazard functions, up and to the right such that ${h_1(t) = \lambda h_0(\lambda t)}$  \textbf{e-f.} Accelerated Failure time models are easily modified to model populations whose survival distributions are shifted (rather than scaled) in respect to time. \textbf{g-h.} The existence of heterogeneity within a population in respect to the risk of death produces a deceleration, or leveling-off of the hazard function and a corresponding long-tail of the survival function. }
\label{fig_3}
\end{SCfigure}

The two most common families of semi-parametric models are Proportional Hazards (PH) models and Accelerated Failure Time (AFT) models.  Proportional hazards models assume interventions alter the hazard function to produce a time-independent proportional change in the risk of death.  In practice, this means that PH models assume that an intervention will produce vertical shift of the hazard function when plotted on log-linear axis populations, just as a change in the Gompertz $a$ or the Weibull $\beta$ parameter does.  (Fig. \ref{fig_3} a-b).  PH models are formally defined by the relation $h_1(t) = \lambda h_0(t)$, where $h_1(t)$ is the hazard function of a population exposed to some intervention with $h_0(t)$ as the control group.  Accelerated failure time models, in contrast to PH models, assume that interventions produce a temporal rescaling of aging, stretching or compressing survival curves such that $S_1(t) = S_0(\lambda t)$ (Fig. \ref{fig_3} c-d).  AFT models offer an intuitive physical interpretation, that interventions extend lifespan by decreasing the rate of the underlying physiological or cell biologic processes that determine the timing of death.  This is equivalent to a change in the Gompertz $b$ or Weibull $\beta$ parameters.

Semi-parametric models are widely used in clinical research and therefore a deep literature exists exploring their behavior in diverse contexts \cite{singer2003applied, kalbfleisch2011statistical}.    Multivariate methods such as Cox regression and Buckley-James regression allow multiple influences of lifespan to be considered simultaneously, for example where environmental factors differ between experimental replicates \cite{lucanic2017impact} or when multiple interventions are applied simultaneously. PH and AFT models often provide very good approximations of empiric data in a variety of organisms, including yeast \cite{liu2018generational}, flies \cite{john2018diet}, nematodes \cite{hughes2016different, stroustrup2016temporal}, and mice \cite{swindel:2009, conti2006transgenic} A variety of approaches exist to identify and compensate for situations where assumptions are not met, including segmenting time or allowing continuously time-varying hazard ratios\cite{hagar2017flexible}.  Additive hazards (AH) models, assuming $h_1(t) = \Delta h+ h_0(t)$ have also been suggested \cite{aalen1993further}. 

AFT models, unlike PH models, produce residual distributions that take the same time units as the death time provided.  This allows residuals to subsequently be used as a time-standardized lifespan distribution for qualitative comparisons between different populations \cite{stroustrup2016temporal} and a convenient means for handling confounding effects of environmental factors \cite{stroustrup:2013}.   In cases where interventions do not slow the aging process, but rather delay it by a fixed interval to produce a rigid shift (rather than a scaling) of the survival curve, interventions can be modeled using similar methods assuming $S_1(t) = S_0(t-\Delta_\tau)$ (Fig. \ref{fig_3} e-f). 

\section{Accounting for heterogeneity within groups with frailty models}
Basic parametric forms like Gompertz and Weibull are often employed with the implicit assumption that all individuals in a population age according to the same parameters. This assumption is rarely justified by experimental data, as even within isogenic populations of laboratory animals housed in controlled laboratory conditions, subpopulations are observed to age in distinctive ways \cite{ pincus2016autofluorescence,tiku2017small,herndon:2002,cannon2017expression,harvanek2016computational} ,\cite{zhang2016extended}.  Heterogeneity is also be produced when individuals respond unevenly to an intervention, as is commonly observed\cite{rea2005stress}, \cite{bensaddek2016micro},\cite{mendenhall2017c}.  

The effect of the heterogeneity on the hazard and survival functions can be accurately modeled even when it cannot be explicitly measured.  This is accomplished by assuming the effect of the unmeasured heterogeneity takes a simple parametric form as it varies between individuals.  The effect is then incorporated in a parametric or semi-parametric model and referred to as a ``frailty'' or ``random effect'' term.   A two-parameter Gompertz model, for example, can be modified to account for heterogeneity in respect to its $a$ parameter by adding a single additional parameter, $\sigma$, to represent the variance of an gamma-distributed $ Z $ random variable, such that the hazard function becomes $h(t) = Z(\sigma) a \exp(t/b)$\cite{aalen:1994}.  In this way, parametric and semi-parametric models can be modified to account for unmeasured heterogeneity by adding a single additional parameter.

Heterogeneity of this kind produces a clear effect on hazard functions--a progressive deceleration of hazard functions relative to the basic underlying parametric form \cite{vaupel1979impact}, leading in some cases to a plateauing (flattening) of the hazard functions at advanced ages (Fig. \ref{fig_3} g-h).  This deceleration arises as a consequence of high-frailty, high-risk subpopulations dying earlier than low-frailty and low-risk sub-populations.  As the high-risk individuals die off, the remaining population increasingly consists of relatively low-risk individuals.  This change in the populations' composition counteracts the increasing risk of each individual, producing a quasi-stationary state in which the hazard rate appears flat \cite{steinsaltz2004markov}. Decelerating hazard functions are observed in most model-organism lifespan data.  In some cases the effects appear subtle, but in many cases late-life deceleration is a dominant feature of empiric hazard functions, limiting the practical application of simple two parameter Gompertz or Weibull models  \cite{vaupel1979impact, vaupel1993compositional, aalen1994effects}, \cite{stroustrup2016temporal}.  Frailty-associated heterogeneity also confounds efforts to experimentally identify the true parametric form of empiric data sets.  Simple parametric models are most easily distinguished by their behavior at late ages, at the tail of the parametric probability distribution.  Heterogeneity masks the underlying form of these tails, undermining biological interpretations that depend on empiric justification of specific parametric forms \cite{stroustrup2016temporal} \cite{vanfleteren:1998}.
\begin{figure}
\centering
\includegraphics[width=0.85\textwidth]{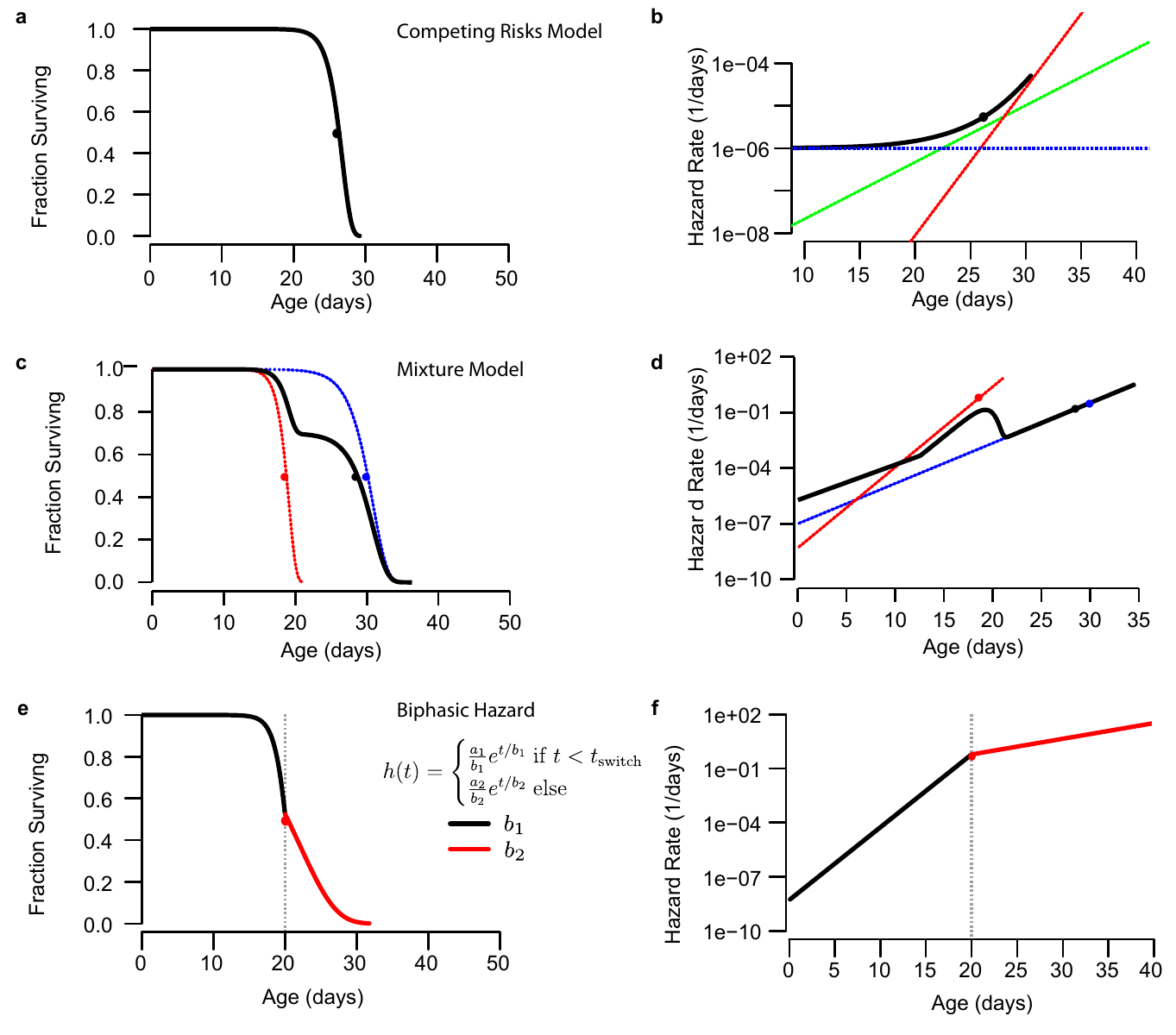}
\caption{\textbf{Modeling different types of  heterogeneity}.  Several techniques exist to describe heterogeneity between individuals within a population and over time during the aging process. \textbf{a-b.} Competing risk models describe the effect that multiple causes of death have on a population's hazard function.  Here, three statistically-independent causes of death exhibit different temporal dynamics.  One cause \emph{(blue)} shows a constant risk over time, another cause \emph{(green)} increases slowly with age and a third \emph{(red)} increases rapidly later in life. The cause-specific hazard functions of each cause sum to produce the all-cause hazar function and survival curve \emph{(black)}. \textbf{c-d.} In come cases, the cause of death of an individual may be predetermined early in life, producing distinct subpopulations dying from distinct causes of death.  Here, thirty percent of individuals  die early according to one cause of death \emph{(red)} and the remainder die according to a second cause \emph{(blue)}.  In this case, the hazard functions do not strictly add. \textbf{e-f.} Survival and hazard functions can be modeled with a ``change-point" that separates distinct phases of aging. Though geometrically compelling, the biological interpretation of segmented hazards is often problematic, as the break-point time must be uncorrelated with each individual's death time. }
\label{fig_4}
\end{figure}

\section{Accounting for multiple causes of death with competing risk models}
	Competing risks models explore the idea that though an organism can die only once, it remains at risk of dying from multiple possible causes up until the moment that one particular cause kills it. These models are most intuitively applied in populations in which individuals are suffering simultaneously from several potentially fatal diseases--for example cancer patients with cardiovascular conditions \cite{albertsen1998competing}.  Competing risk models explore quantitative relationships between causes of death, and provide a framework in which to disentangle the effects of interventions that may not act homogeneously across all mechanisms driving aging.  For example, competing risk Cox regression and AFT regression models can evaluate the simultaneous influence of multiple interventions on multiple causes of death\cite{austin2016introduction}.  

Two classes of competing risks differ in their interpretation.  One interpretation assumes that all individuals share the same risk for each cause of death, and therefore describes the qualitatively different outcomes produced by physiological changes experienced in the same way by all individuals. Another interpretation assumes risks differ across sub-populations, with the extreme case being that certain individuals have their cause of death pre-determined early in life.  This latter assumption often called ``Mixture models" (Fig \ref{fig_4}c-d), is of particular contemporary interest given the many types of distinct subpopulations now being identified within isogenic populations \cite{zhao2017twophasegems}$^{\bullet\bullet}$\cite{vaupel1985heterogeneity, suda2009analyzing, shi2017mating}. 

Competing risk models involve data where each death time $T_i$ is paired with a label $C_i$ describing the cause of death.  This allows an intuitive decomposition of the all-cause hazard into the sum of several  cause-specific hazard functions, \\ ${h(t) = \sum_i h_i(t) = \sum_i \lim P(t \leq T+\Delta t | T > t, C = C_i )}$ as  $\Delta t \rightarrow 0 $ \cite{kalbfleisch2011statistical}, shown in Figures \ref{fig_1} and \ref{fig_4}a-b.

Competing risk models were originally developed as an analytic tool for interpreting lifespan data.  However, the assumptions required to accomplish this task provide a useful starting point for experimental researchers attempting to understand the physiological basis of aging.  Are different causes of death influenced equally by interventions in aging?  Can genetically defined aging pathways be contextualized in respect to their mutual dependence or independence in influencing different causes of death?  These questions can be rendered empirically testable by rephrasing them as testable quantitative predictions of competing risk models.  Recent work \cite{stroustrup2016temporal} \cite{zhao2017twophasegems},\cite{galimov2018coupling},\cite{leiser2016age} demonstrates the promise of competing risk models to describe aging phenomena working at longer timescales, where different physiological changes throughout life interact to determine the timing of death and the shape of hazard functions and survival curves. 

\section{Describing multi-stage aging processes}
The physiological changes driving aging may vary over the course of an individual's lifespan.  For example, recent work has identified two phases of aging distinguished according to changes in movement and behavior, transitions in gut permeability \cite{dambroise2016two}, changes in rate of apparent morphologic changes\cite{zhang2016extended}, transcriptomic changes, \cite{eckley2017transcriptome,jovic2017temporal} ,\cite{angeles2017caenorhabditis}, protein aggregation \cite{ben2009collapse,david2010widespread}, and differences in susceptibility to early-life \cite{zhao2017twophasegems} or late-life \cite{podshivalova2017mutation} bacterial infection.  The best quantitative approach to model these phases remains an open question and active area of research.  In some cases, the physiological distinctions between some phases may not correspond to distinct phases of the hazard function, in which case no substantial changes are needed in survival analysis.  In other cases, individuals in each state may exhibit distinctive risk of death, and a class of multi-state models often called ''illness-death`` models will be applicable \cite{andersen2012interpretability}.

\section{Biphasic and segmented hazard functions}
A class of models have been proposed that separate aging into distinct phases relative to some landmark, or "change-point" specified in chronological time \cite{noura1990segmented,johnson2001age, michalski2001heating,baeriswyl2010modulation}. (Fig \ref{fig_4}.e-f).  The biological interpretation of these models is often problematic. For the hazard function to exhibit distinct phases, the change-point between them  must be determined either by some internal physiological process or by some external factor--e.g. a shift in environment or diet.  In the latter case, it is the environment that is biphasic, not aging itself.  The former case involves an apparent contraction, in that individuals variable in their lifespan must nevertheless synchronously switch between phases. This would require that the physiologic processes specifying the change-point in all individuals be at most weakly correlated with the physiologic processes determining the timing of death of each individual.  A more physiologically plausible interpretation would seem to be that biphasic hazard functions arise not from a biphasic aging process but instead from the distinct contributions of unidentified subpopulations, as described by competing risks and mixture models.

Biphasic hazard models in many cases may simply represent an over-fitting or mis-fitting of empiric data. For example, an apparently biphasic Gompertzian hazard function may be better explained by a single-phase inverse Gaussian distribution or a single-phase Gompertzian distribution with an extra parameter correcting for the effects of frailty.

\section{Summary}
Aging research is undergoing a period of rapid discovery and characterization of genetic, pharmaceutical, and dietary interventions in aging.  Several of these therapies are being explored for translational potential, and lifespan data from human clinical trials may soon be available in which patients' survival is altered by molecular perturbation of basic aging processes.  This growing abundance of lifespan data demands the thoughtful application of statistical methods. 

For new projects, familiar analytic techniques should be re-evaluated.  In particular, experimentalists should recognize that the Gompertz hazard parameterization became standard decades ago, long before the high-resolution data needed to validate it became available.  Experimentalists should, as an alternative, consider frailty-corrected Gompertz distributions or semi-parametric methods like AFT or PH regression. Where parametric analyses are justified, best-practice maximum likelihood fitting methods should be employed. Finally, competing risks and mixture models should seriously considered in situations where multiple aging processes may influence one or more outcomes in aging. Employing this broad tool set of analytic approaches, experimentalists can move beyond humble significance testing to instead use lifespan data as a versatile means for studying the physiological dynamics of aging.

\section{Acknowledgments}
I thank members of my group as well as Javier Apfeld for stimulating discussions and feedback on the manuscript.  I further acknowledge the support of the Spanish Ministry of Economy, Industry and Competitiveness (MEIC) to the EMBL partnership, the Centro de Excelencia Severo Ochoa, and the CERCA Programme / Generalitat de Catalunya.  This work was supported by the MEIC Excelencia award BFU2017-88615-P, and by an award from the Glenn Foundation for Medical Research.

\section{Work Cited}

\bibliographystyle{elsarticle-num-names}

\bibliography{competing_risks_review_merged_references}

\end{document}